# Portals and Task Innovation: A Theoretical Framework Founded on Business Intelligence Thinking

**Dima Jalal**
MBA/Quality Management Candidate
The University of Jordan
Faculty of Business

**Mutaz M. Al-Debei**
Assistant Professor
Director of the University of Jordan Website Office
The University of Jordan, Faculty of Business
Department of Management Information Systems
m.aldebei@ju.edu.jo

## Abstract

The main aim of this study is to develop a theoretical framework for the success of Web portals in promoting task innovation. This is deemed significant as yet little research has tackled this important domain from the business intelligence perspective. The D&M IS Success Model was used as a foundational theory and then was refined to match the context of the current research. Importantly, in this study, system quality and information quality constructs were defined on the basis of portals' characteristics since a mapping was conducted between the most significant functions and features of Web portals and quality constructs. The developed framework is deemed useful for theory and practice. From theoretical perspective, the dimensions that affect the perceived quality of Web portals are identified, and the measures that affect each quality dimension are also defined. On the practical level, contributions gained by this study can be observed in terms of the benefits decision makers, strategists, operational employees and IT developers can gain. Assessing portals success in improving task innovation is important to help managers (i.e. decision makers) in making appropriate decisions concerning the adoption of portals' technology, by weighing its benefits against the costs needed to establish and run such a technology. Moreover, assessing Web portals' success gives some insight to IT developers and designers concerning what aspects should be taken when designing and establishing high quality portals, and what functions and features should be contained that would affect the perceived quality of portals and therefore users' intention to use portals.

## 1. Introduction

Portals are considered to be a type of information systems used to gather, manage, share, and utilize information that has been stored in disparate databases throughout the

١

organization (Aneja et al., 2000; Bock, 2001; Brabston & McNamara, 1998; McManis et al., 2001; Yu et al., 2001); thus portals provide users with a single point of access to personalized information needed to make informed business decisions. Portals can bring significant benefits to organizations at both the individual and organizational levels. This is because portals enjoy the ability to integrate disparate information sources and allow easier access to existing applications within the organization; and hence portals allow staff to find the information and knowledge that they need to do their jobs effectively (Collins, 2002: Detlor, 2004; Terra and Gordon, 2003). Moreover, portals can further enhance effectiveness by supporting communication between individuals and workgroups; thus allowing increased collaboration (Benbya et al, 2004; Detlor, 2000; Dias, 2001). In addition to that, portals can improve internal operations and collaboration with external business partners, such as customers and suppliers (Dias, 2001; Detlor, 2000). Besides, portals can reduce information overload, reduce organizational costs, and enhance employee innovation and business intelligence capabilities (Tojib et al., 2006). Portals also benefit in streamlining business processes, increasing efficiency and productivity, and improving employee satisfaction due to increased convenience in accessing relevant applications and information. All that due to the greater collaboration and learning opportunities provided through portals (Rahim and Singh, 2006)

Today, many organizations especially large ones, apply portals and use them as part of their working procedures. Despite the restricted IT budgets of many organizations, investments in portal solutions are still growing. Portal projects are usually considered as complex, time and cost-consuming, with a high failure risk (Remus, 2006), but some organizations are still investing huge amounts of money in building, establishing and running portals, without assessing the actual benefits of their portal implementations (Brown et al., 2007). In fact, portals success in delivering intended benefits is dependent on the degree to which people accept portals' technology and intend to use it on a continual basis. People will not tend to use portals if they are not satisfied with them, or do not perceive that portals are useful to them. Users' perception of portals' quality and their satisfaction will lead to future use, and therefore success of portals in delivering intended benefits such as task innovation and business intelligence.

This paper aims at developing a portal success model that explains how web portals can be effective in delivering intended benefits at the individual level mainly by promoting individual task innovation. Therefore, success of portals will be measured in terms of their impact on task innovation, from the perspective of business intelligence. A framework is developed after examining IS Success Models and choosing the DeLone and McLean IS Success Model (2003) as the basis to develop a portal success model.

Little research exists on assessing the success of Web portals in developing countries, although IS Success Models, mainly DeLone and McLean IS Success Model, were used to assess the success some systems such as government to citizen (G2C) eGovernment systems (Wang and Liao, 2008), E-Commerce (Molla and Licker, 2001; Zhu and Kraemer, 2005), Knowledge management systems (Wu and Wang, 2006; Kulkarni et al. 2006), Web-based applications (Kwan, 2006), and Portals (Busaidi, 2010; Urbach et al., 2010). So, this paper addresses this gap by developing a portal success model that is suitable to assess portals success. Importantly, this paper makes a contribution to the literature since the most significant portals' functions and features are identified and related to the quality dimensions of portals, mainly system quality and information quality, to which they contribute.

## 2. Literature Review and Theoretical background

### 2.1 Overview on Web Portals

A portal can be defined as a single point of access (SPOA) for the pooling, organizing, interacting, and distributing of organizational knowledge and creating business intelligence (Bock, 2001; Kendler, 2000; Schroeder, 2000). Therefore, portals synchronize knowledge and applications, creating a single view into the organization's intellectual capital (Benbya et al., 2004). Portals' competitive advantage depends on their abilities to filter, target, and categorize information so that users will get only what they need (Eckel, 2000). By getting customized information, users are more able to make informed business decisions and more able to be innovative in performing their tasks.

As portals are considered as business intelligence systems that help in the retrieval of accurate and timely information about financial operations, customers, and products, in order to gain analytical insight into business problems and opportunities, business organizations are continually using portals to support business analysis and decision making (Gangadharan and Swami, 2005).

Portals have evolved from simple providers of Web page access and corporate databases to support intelligent management, integration of applications and collaborative processing. Portals can be considered as an evolution of data warehouses extending its application to Intranet giving access to all information resources and knowledge of a firm (Dias, 2001). The advantage of portals is their ability to integrate and personalize several technologies (e.g. groupware, databases, data warehouses, e-mail, meta-data, intelligent

٣

management systems, and other technologies) in a unique business management tool. Portals assist users in business intelligence processes and encourage them to take more informative decisions and to be more innovative.

The primary objective for developing portals may vary from one organization to another (Hazra, 2002), although in general it is to create a working environment where users can easily navigate to find the information they specifically need to perform their operational or strategic functions quickly and to make informed decisions (Collins, 2001).

### 2.1.1 Web Portals: Functions and Features

Many research have been done in an attempt to identify portals' functions and features, which are fairly difficult to define separately as they unite inter-related components (Raol et al, 2003). Functions are the components that provide access to the range of disparate enterprise databases and information resources and the ease with which users can set up personalized access to enterprise and external information resources (White, 2000). In most portals, these functions and features may include, but are not limited to, security (Benbya et al, 2004; Raol et al, 2003; Collins, 2001), customization and personalization (Dias, 2001; Kotorov and Hsu, 2001; Raol et al, 2002; Collins, 2001; Benbya et al, 2004), content management (Dias, 2001; Raol et al, 2002; Benbya et al, 2004; Collins, 2001), Taxonomy (Benbya et al, 2004; Collins, 2001), integration (Benbya et al, 2004; Dias, 2001; Collins, 2001; Detlor, 2000; Mack et al, 2001), searchability (Detlor, 2000; Dias, 2001; Raol et al, 2002; Kotorov and Hsu, 2001; Mack et al, 2001; Benbya et al, 2004; Collins, 2001), scalability (Benbya et al, 2004; Raol et al, 2003), collaboration (Benbya et al, 2004; Raol et al, 2003; Collins, 2001; Mack et al, 2001), presentation (Collins, 2001), administrative tools (Collins, 2001), profiling (Benbya et al, 2004), accessibility (Breu and Hemingway, 2001; Mack et al, 2001), and eas to use (Raol et al, 2003; Dias, 2001; Benbya et al, 2004).

Based on synthesizing the related literature, the main functions and features of portals are defined as: **(1) Content Management and Tailorability** which provides users with the ability to adjust and tailor accessed data based on users' specific requirements and preferences, and this function encapsulates Customization/ Personalization/ Profiling/ Content Management/ Taxonomy/ Presentation**;** **(2) Integration** which aims at bringing, harmonizing and synchronizing data existing in different formats in incompatible applications all together, and then presenting it on a unified interface (i.e. the portal); **(3) Security** which provide users with a secure access to diverse range of resources, by describing the levels of access each user or groups of users are allowed for each portal application and software function included in the portal; **(4) Searchability** which allow users to retrieve required information directly by using search engines, instead of browsing through the different information categories; **(5) Collaboration** which provide users with collaborative tools needed to enforce and optimize work and process collaboration inside and outside the organization; **(6) Scalability** which describes the capability of the system to cope and perform under an increased or expanding workload; and **(7) Accessibility** which describes the ability to access the system from anywhere at anytime.

We postulate, based on analyzing the literature, that employing portals with the functions and features above supports business intelligence and task innovation within organizations. This is because functions as "Content Management and Tailorability" and "Searchability" help in retrieving and tailoring portals' information content according to users' needs, which would help them to get the needed information in the needed format that makes it possible to analyze data and produce reports on which they can base their decisions. Users of portals can also gain more insight into business problems and related issues that make them more innovative. "Integration" feature helps in getting data existing in different applications and presenting it in an integrated and harmonized manner. As a result users would get the information needed faster in a way that gives them the opportunity to be more able to make more informed decisions. Collaborative tools allow users to get closer to colleagues that make it easier to communicate internally so as to get more insight into business issues and problems, to collaborate in order to optimize business processes and to come up with innovative solutions to business problems. Also, collaboration tools help in getting closer to customers, and in being more able to understand their needs and requirements, which is necessary for making informed strategic decisions about how to deal with customers and how to fulfill their needs and solve their problems in an innovative manner. "Accessibility" makes it possible for users to access and navigate portals from wherever they are; this encourages users to work at their convenience and be more innovative when they feel that they can get the information they need easily and effectively, in a way that makes it easier for them to perform their tasks more innovatively. Also, "Security" allows a secured access to information databases, which encourages users to use portals as a tool that provides them with needed information to perform tasks and make informed decisions.

## 2.2 Background Theory: IS Success Model

Web portals are considered as a type of knowledge management systems that provide access to integrated applications and databases, they act as a business intelligence tool that supports the decision making process and sometimes fosters task innovation. Many models were used to measure the success of information and knowledge management systems, such as DeLone and McLean IS Success Model, which has proven some validity in assessing the success of knowledge management systems. Therefore, DeLone and McLean IS Success Model will be used as a foundation to build a theoretical framework adequate for measuring portals' success in promoting task innovation

Looking at portals as a type of information systems, it is not easy to define their success, since there are several definitions and measures of IS success provided in the IS literature. This is because there are different stakeholders who assess IS success in an organization (Grover et al., 1996), and each group assesses success from its perspective. Furthermore, IS success also depends on the type of system being evaluated (Seddon et al., 1999).

In order to provide a more general and comprehensive definition of IS success, one that covers these different perspectives, DeLone and McLean (1992) reviewed the existing definitions of IS success and their corresponding measures, and classified them into six major variables: "System Quality", "Information Quality", "Use", "User Satisfaction", "Individual Impact", and "Organizational Impact". They then created a multidimensional measuring model with interdependencies between the different success variables, which became very popular.

Several studies have been conducted in an attempt to extend or re-specify DeLone and McLean original model (1992). For example, some researchers either suggested that further dimensions should be included in the model, or they presented alternative success models (Seddon, 1997; Seddon and Kiew, 1994). Others have focused on the model's application and validation (i.e. Rai et al., 2002) since DeLone and McLean have called for that. Conducted research (i.e. Seddon, 1997; Pitt et al., 1995) have raised some critics and weaknesses of the original model (1992), to which DeLone and McLean have responded and developed an updated model in (Delone and Mclean, 2003). The updated model consists of six interrelated dimensions of IS success: "Information Quality", "System Quality", "Service Quality", "Intention to Use", "Use", "User Satisfaction", and "Net Benefits". The model can be interpreted as follows: A system can be evaluated in terms of the information, system, and service quality; these characteristics affect subsequent use or intention to use, and user satisfaction. As a result of using the system, certain benefits will be achieved. If "Net Benefits" are positive from the perspective of the owner or sponsor of the system, then the system will be re-used continually, thus influencing and reinforcing subsequent "Use" and "User satisfaction". These feedback loops are still valid, even if the "Net Benefits" are negative. The lack of positive benefits is likely to lead to decreased use and possible discontinuance of the system or of the IS department itself (e.g. outsourcing). The updated model is shown in Figure 1.

The key modifications in the updated model (2003) are summarized in the following:
- The inclusion of "Service Quality" as an additional aspect of IS success
- The elimination of "Individual Impact" and "Organizational Impact" as separate variables, and their replacement with "Net Benefits".
- The clarification of the "Use" construct, by measuring "Intention to Use" (an attitude) rather than "Use" (a behavior)

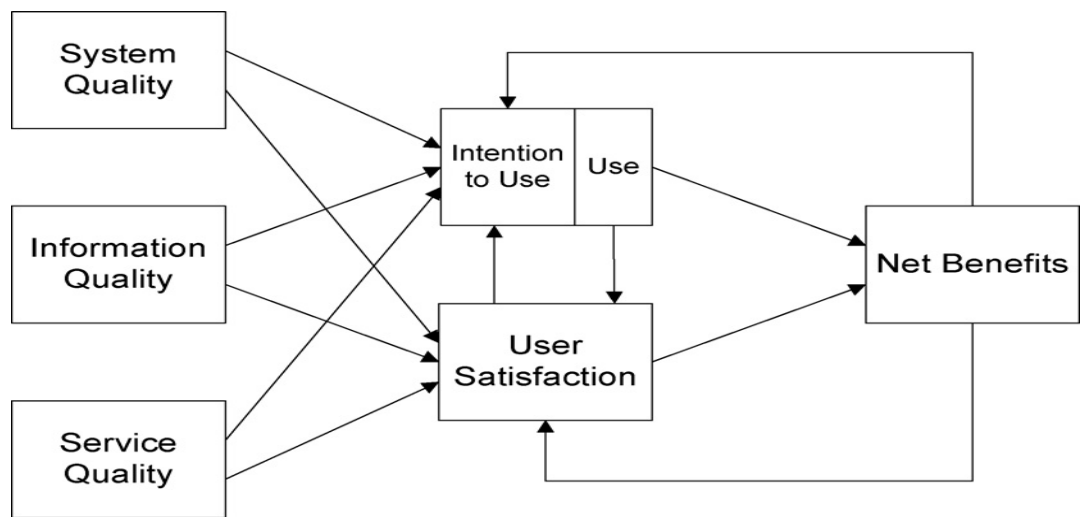

**Figure 1. Updated D&M IS Success Model** (Delone and Mclean, 2003)

## 2.3 Measuring the Success of Web Portals

Existing measurement approaches to assess portal success in practice usually utilize monetary indicators, such as return on investment (ROI) or total cost of ownership (TCO) and other cost-benefit analysis methods (White, 2003). To arrive at a comprehensive measurement of portals' success, we need to consider both the tangible and intangible effects of a portal to detect potential improvements, and to justify present and future investments in portal solutions, for example, portals' success cannot be measured solely by its reach, one should not purely rely on "hit counts" as a measure of success (Damsgaard and Scheepers, 1999).

Successful and effective portals are those that are used repeatedly by its users on a frequent basis for extended periods of use, and this cannot be achieved without users being satisfied about portals' quality and service that they become willing to use/ continue using portals. If firms need to build successful quality portals, they have to take into account the dimensions that affect portals quality. Moraga et al. (2004) defined these dimensions in their model, known as the Portal Quality Model (PQM), that they built based on the SERVQUAL model that was proposed by Parasuraman et al. (1998) and the GQM (Goal Question Metric) method that was proposed by Basili et al. (1994). In this model, the different dimensions of the SERVQUAL model to the portal context were adapted and some of them were split up into sub-dimensions, in order to create a more specific model.

Focusing on the user-perceived service quality of Web portals, Yang et al. (2005) developed and validated a five-dimension service quality instrument involving: "Usability", "Usefulness of Content", "Adequacy of information", "Accessibility", and "Interaction". This scale provides a useful instrument for researchers aiming to measure the service quality of Web portals and for portal managers who wish to improve their service performance. In order to measure user satisfaction with employee portals, Sugianto and Tojib (2006) proposed using the B2E Portal User Satisfaction (B2EPUS) model, which is based on the End-User Computing Satisfaction measure (EUCS) developed by Doll and Torkzadeh (1988). They identified nine dimensions of the B2E portal "User Satisfaction", which are: "Information Content", "Ease of Use", "Convenience of Access", "Timeliness", "Efficiency", "Security", "Confidentiality", "Communication", and "Layout". This scale can also be adapted or supplemented to fit specific needs. For example, in cases where an organization wishes to measure the extent to which the portal delivers the intended benefits, or when an organization plans to create a workplace that is conducive to and supports employee satisfaction and productivity.

Masrek (2007) has proposed another approach to assessing user satisfaction with campus portals, which is based on an extract of the updated D&M IS Success Model (DeLone and McLean, 2003). Masrek (2007) aimed at evaluating the effectiveness or success of campus portal implementation from the perspective of students as users, and sought to investigate the influence of individual factors comprising attitudes towards the portal, personal innovativeness and Web self-efficacy on the effectiveness of the portal. The study found that of the three predictors investigated, only attitudes towards the portal were found to be significantly correlated with IS effectiveness dimensions. Urbach et al. (2010) introduced a theoretical model to gain a better understanding of employee portal success that is based on the original DeLone and McLean IS Success Model (Delone and Mclean, 1992). They tested the associations between different models' success dimensions and found that besides the factors contributing to IS success in general, other success dimensions – like the quality of the collaboration and process support – have to be considered when aiming for a successful employee portal.

Based on related literature and mainly on the studies presented in this section, we, in the following section, propose a theoretical framework for assessing the success of Web portals from task innovation and business intelligence perspective.

## 2.4 Proposed Theoretical Framework and Hypotheses Development

DeLone and McLean IS Success Model was used by many studies to evaluate the success of various types of information systems, such as government to citizen (G2C) eGovernment systems (Wang and Liao, 2008), E-Commerce (Molla and Licker, 2001; Zhu and Kraemer, 2005), Knowledge management systems (Wu and Wang, 2006; Kulkarni et al. 2006), Web-based applications (Kwan, 2006), and Portals (Busaidi, 2010; Urbach et al., 2010). Yet, there is little research on the use of DeLone and McLean IS Success Model to assess Portals' success from business intelligence perspective. Hence, this paper is devoted to develop a theoretical framework based on DeLone and McLean IS Success Model (Delone and Mclean, 2003) to form the foundation for assessing the success of Web portals in promoting task innovation.

The updated DeLone and McLean IS Success Model (2003) accounted for benefits occurring at any level of analysis (i.e. individual or organizational), the choice of what kind of impact (individual or organizational impact) to be measured depends on the systems being evaluated and their purposes. Here, we are interested in examining the benefits that an individual can gain from the use of Web portals, mainly the benefit of promoting task innovation. Therefore, the construct "Net benefits" will be replaced by "Task innovation". The proposed theoretical framework that will be used to assess portals' success in promoting task innovation is the one shown in Figure 2.

"Service Quality" will be excluded from the three quality dimensions of Web portals, since we are looking at portals as a business intelligence and knowledge management system whose quality primarily depends on the design characteristics and features, functioning capabilities and the quality of information provided. Moreover, "Service Quality" construct can be considered merely a subset of systems quality (Yang et al. 2005). Wu and Wang (2006) supported knowledge management system success based on DeLone and McLean IS Success Model while excluding service quality as a quality dimension. Also, the three hypotheses that were associated with the Service Quality construct were not found to be significant (unsupported) in the study of Petter and McLean (2009).

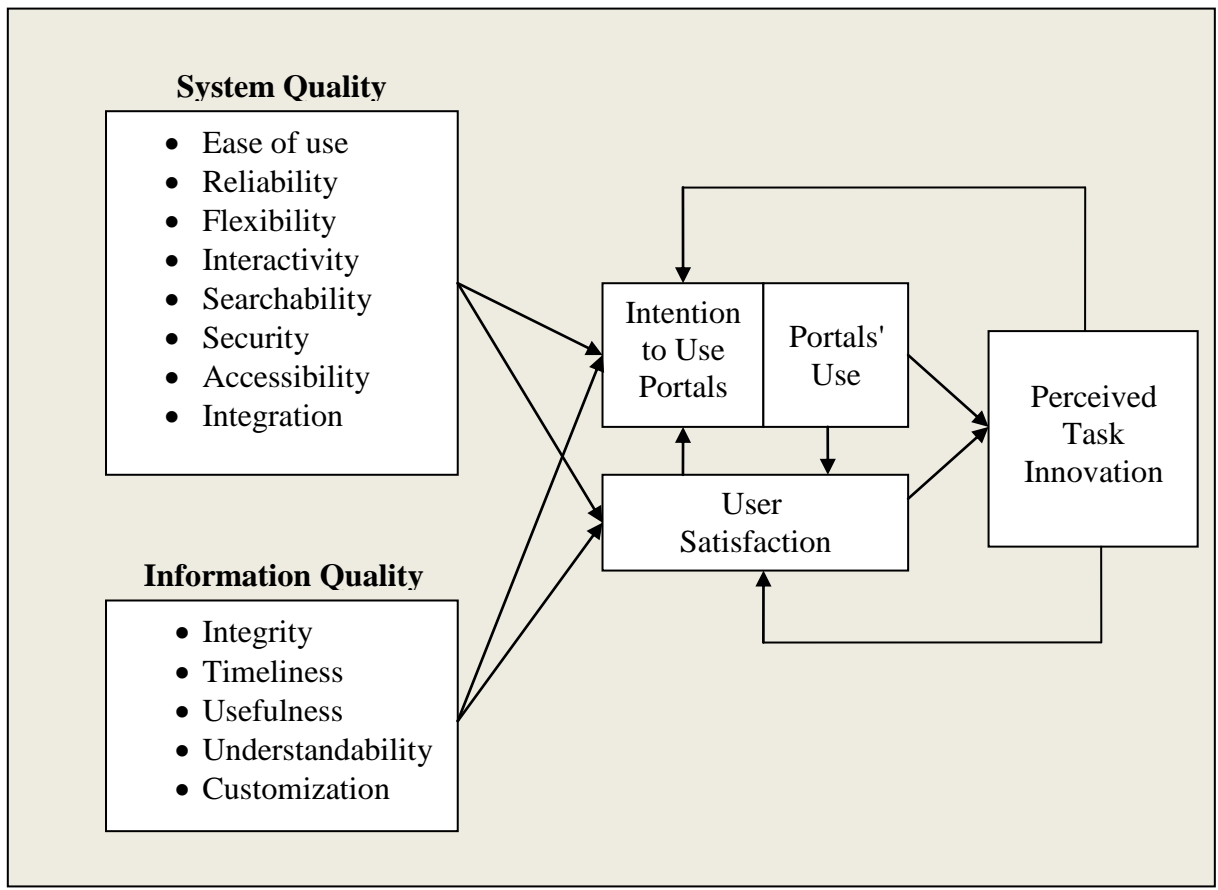

**Figure 2. Proposed Theoretical Framework for Web Portals' Success (WPS)**

## 2.4.1 Constructs of Web Portals' Success (WPS) Theoretical Framework

### 2.4.1.1 System Quality and Information Quality

In this paper, an attempt is done to relate functions and features of portals to the quality dimensions to which they contribute. To achieve this objective, one should be able to distinguish each dimension of the quality dimensions is related to what component of the Information Systems (i.e. portals). For this purpose, a brief explanation would clarify the main components of Information Systems. Any information system depends on the resources of hardware (machines and media), software (programs and procedures), networks (communications media and network support), data (data and knowledge basis), and people (end users and IS specialists), to perform input, processing, output, storage, and control activities that convert data resources into information products.

The "System Quality" Construct represents the quality of the information system itself, and it is a measure of the extent to which the system is technically sound (Gorla et al., 2010). In other words, the system quality can be determined by the quality of its hardware or software components that represent data and information capturing, processing, storage and retrieval capabilities.

"Information Quality" refers to the quality of outputs the information system produces (DeLone and McLean, 1992), which can be in the form of reports or online screens.

Based on the discussion above, all functions and features that are related to hardware and software components will be considered to have an impact on "System Quality", and those functions that serve in managing data and information will be considered to have an impact on "Information Quality". Based on this, "Content Management and Tailorability" is affecting Information Quality, since this function allows the user to be provided with an appealing, personalized and understandable visual appearance. Other functions and features such as: "Integration", "Search Capabilities", "Collaboration", "Security", "Scalability", "Accessibility" and "Ease of use" are related to "System Quality", since they need specific software tools or applications.

### 2.4.1.1.1 Web Portal's System Quality

Petter et al (2008) defined "System Quality" as "The desirable characteristics of an information system". Seddon (1997) notes that ''system quality is concerned with whether there are bugs in the system, the consistency of user interface, ease of use, quality of documentation, and sometimes, quality and maintainability of program code" (p. 246).

In this paper, "System Quality" will be referred to as "Portal's System Quality" and it will be measured by the following dimensions: "Ease of use", "Reliability", "Flexibility", "Interactivity", "Searchability", "Security", "Accessibility", and "Integration".

### 2.4.1.1.1.1 The Relationship between Portal's System Quality and Intention to Use

There is mixed support for the relationship between "System Quality" and "Intention to Use" at the individual level of analysis within the literature. Some researchers have found that perceived ease of use is not significantly related to intention to use (Subramanian, 1994; Agarwal & Prasad, 1997; Lucas & Spitler, 1999; McGill et al., 2003; Klein, 2007), on the other hand many studies that measure system quality as perceived ease of use have found positive relationships with behavioral intentions to use the system (Venkatesh & Davis, 2000; Venkatesh & Morris, 2000; Hong et al., 2001/2002). Thus, it will be assumed that higher portal's system quality will lead to higher intention to use the portal.

*H1: Portal's System Quality positively influences Intention to Use*

### 2.4.1.1.1.2 The Relationship between Portal's System Quality and User Satisfaction

At the individual unit of analysis, there is strong support for the relationship between "System Quality" and "User Satisfaction" (Iivari, 2005). For knowledge management systems, system quality was found to be strongly related to user satisfaction (Kulkarni et al., 2006; Wu & Wang, 2006; Halawi et al., 2007). For Web sites, system quality, measured as reliability and download time, is significantly related to user satisfaction in two different studies (Kim et al., 2002; Palmer, 2002). System quality, measured in terms of perceived ease of use, also has a significant relationship with user satisfaction (Devaraj et al., 2002; Hsieh & Wang, 2007). Thus, it will be assumed that higher system quality will lead to higher user satisfaction.

*H2: Portal's System Quality positively influences Users' Satisfaction with portals*

### 2.4.1.1.2 Information Quality

Petter et al (2008) defined "Information Quality" as "The desirable characteristics of an information system's outputs". "Information Quality" can be measured by the following dimensions: "Integrity", "Timeliness", "Usefulness", "Understandability" and "Customization"

### 2.4.1.1.2.1 The Relationship between Information Quality and Intention to Use

Few studies have examined the relationship between "Information Quality" and "Intention to use" at both the individual and organizational levels. One reason for this is that information quality tends to be measured as a component of user satisfaction measures, rather than being evaluated as a separate construct. Two studies found that information quality is not significantly related to intention to use (McGill et al., 2003; Iivari, 2005), whereas another study of knowledge management systems found that information (or knowledge) quality was significantly related to intention to use (Halawi et al., 2007). Thus, it will be assumed that higher information quality will lead to higher intention to use.

*H3: Information Quality of portals positively influences Intention to Use*

### 4.2.1.1.2.2 The Relationship between Information Quality and User Satisfaction

The relationship between "Information Quality" and "User Satisfaction" is strongly supported in the literature (Iivari, 2005; Wu & Wang, 2006). Studies have found a consistent relationship between information quality and user satisfaction at the individual unit of analysis (Seddon & Yip, 1992; Seddon & Kiew, 1996; Bharati, 2002; Rai et al., 2002; McGill et al., 2003; Almutairi & Subramanian, 2005; Wixom & Todd, 2005; Kulkarni et al., 2006; Chiu et al., 2007; Halawi et al., 2007). Thus, it will be assumed that higher information quality will lead to higher users' satisfaction.

*H4: Information Quality of portals positively influences Users' Satisfaction with portals*

### 2.4.1.2 User Satisfaction

User satisfaction has been defined as 'the extent to which users believe that the information system available to them meets their information requirement' (Ives, Olson, & Baroudi, 1983: 785).

User satisfaction is the affective attitude to a portal of an employee who interacts directly with it (Doll and Torkzadeh, 1988; Tojib et al., 2006). It refers to the feeling of pleasure or displeasure that results from aggregating all the benefits that a person hopes to receive from interaction with the IS (portals in this case) (Masrek et al. 2009). User satisfaction reflects a user's perceptions of both quality of the system itself and the quality of the information that can be obtained from it. (McGill et al. 2003).

It was found that there is a significant relationship between user satisfaction and intention to use (Wixom & Todd, 2005; Halawi et al., 2007). Besides, satisfaction was found to be a major determinant of continued usage (Igbaria and Tan, 1997; Bokhari, 2005). Arguably, an individual being satisfied with a technology after the initial trial may have a high intention to continue using the technology. This is due to the positive reinforcement of the attitude toward the technology after using it. Thus, it will be assumed that higher users' satisfaction leads to higher intention to use.

*H5: Users' Satisfaction with portals positively influences Intention to Use*

Although Yuthas & Young (1998) found that user satisfaction was only weakly correlated with decision making performance, but most empirical results have shown a strong association between user satisfaction and system benefits (Iivari, 2005). User satisfaction has been found to have a positive impact on a user's job (Yoon & Guimaraes, 1995; Guimaraes & Igbaria, 1997; Torkzadeh & Doll, 1999), to improve performance (McGill et al., 2003), to increase productivity and effectiveness (Igbaria & Tan, 1997; Rai et al., 2002; McGill & Klobas, 2005; Halawi et al., 2007), to improve decision making (Vlahos & Ferratt, 1995; Vlahos et al., 2004), and to enhance job satisfaction (Morris et al., 2002). Thus, it will be assumed that higher users' satisfaction leads to greater task innovation.

*H6: Users' Satisfaction with portals positively influences Task Innovation*

### 2.4.1.3 Intention to Use, and Portals' Use

The construct of "Intention to Use" is a measure of the likelihood a person will employ an application. It is a predictive variable for system use. (Wu and Wang, 2006). Intention of people to adopt and use a specific technology can be explained by the technology acceptance model (TAM) that was developed by Fred Davis in 1985. TAM examines the mediating role of perceived ease of use and perceived usefulness in their relation between systems characteristics (external variables) and the probability of system use (an indicator of system success). On the other hand, portal's "Use" construct measures the perceived actual use of the portal by an organization's staff. Petter et al (2008) defined "System Use" as "The degree and manner in which staff and customers utilize the capabilities of an information system". Burton-Jones and Straub (2006) defined individual-level system usage as an individual user's employment of one or more features of a system to perform tasks.

Little research has examined the relationship between use and user satisfaction, while the reverse relationship, between user satisfaction and use, is more examined in previous research, so additional research is needed to evaluate this relationship. Seddon & Kiew (1996) found that, in a mandatory context, use was not related to user satisfaction, whereas this relationship was found to be significant in other studies (i.e. Guimaraes et al., 1996; Chiu et al., 2007; Iivari, 2005). Thus, it will be assumed that there is a positive relationship between use and user satisfaction.

> ***H7: Positive experience of Portals' Use positively influences Users' Satisfaction with portals.***

As for the relationship between system use and benefits at the individual level, empirical studies provide moderate support. Some studies suggest that there is no relationship between use and net benefits (Lucas & Spitler, 1999; Iivari, 2005; Wu & Wang, 2006). On the other hand, other studies have found that IS use is positively associated with improved decision making. Yuthas & Young (1998) found that the duration of system use is correlated with decision performance. Burton-Jones & Straub (2006) found a strongly significant relation between system usage and task performance. Many studies confirm significant relationships and/or correlations between system use and net benefits (Goodhue & Thompson, 1995; Yoon & Guimaraes, 1995; Seddon & Kiew, 1996; Abdul-Gader, 1997; Guimaraes & Igbaria, 1997; Igbaria & Tan, 1997; Torkzadeh & Doll, 1999; Weill & Vitale, 1999; D'Ambra & Rice, 2001; Rai et al., 2002; Almutairi & Subramanian, 2005; Kositanurit et al., 2006). Thus, it will be assumed that there is a relationship between use and task innovation

- ***H8: Positive experience of Portals' Use positively influences Task Innovation***

### 2.4.1.4 Task Innovation

As it was mentioned earlier, we are interested in examining the benefits that an individual can gain from the use of Web portals. Some of the individual benefits that can be gained by employees through the use of the portal are: task performance, job efficiency, and overall usefulness (Davis, 1989). Torkzadeh and Doll (1999) developed a four-factor, 12- item instrument for measuring the individual impact of IS. They identified the following individual impact dimensions:

- ❖ Task productivity—the extent to which an application improves the user's output per unit of time;
- ❖ Task innovation—the extent to which an application helps users create and try out new ideas in their work;
- ❖ Customer satisfaction—the extent to which an application helps the user create value for the firm's internal or external customers; and
- ❖ Management control—the extent to which the application helps to regulate work processes and performance.

In this paper, the major concern is to examine the success of Web portals in promoting task innovation, therefore the focus will be on the construct "Task Innovation" that is one of the dimensions of individual impact of IS identified by Torkzadeh and Doll (1999).

When measuring net benefits, using perceived usefulness as the metric, many studies have found a relationship between use of a system and behavioral intention (Subramanian, 1994; Agarwal & Prasad, 1999; Venkatesh & Morris, 2000; Hong et al., 2001/2002; Chau & Hu, 2002; Malhotra & Galletta, 2005; Wixom & Todd, 2005; Klein, 2007). Halawi et al. (2007) identified a significant relationship between intention to use and net benefits as measured by improvements in job performance. Thus, it will be assumed that greater task innovation will lead to greater intention to use.

- *H9: Greater Task innovation positively influences Intention to Use*

Several studies (Seddon & Kiew, 1996; Devaraj et al., 2002; Rai et al., 2002; Kulkarni et al., 2006; Hsieh & Wang, 2007) have found a positive, significant relationship between perceived usefulness (i.e., net benefits) and user satisfaction. Three studies found that the impact an expert system has on a user's job directly affects user satisfaction (Yoon et al., 1995; Guimaraes et al., 1996; Wu & Wang, 2006). Abdul-Gader (1997) found a significant correlation between perceived productivity and user satisfaction of computer-mediated communication systems in Saudi Arabia. Thus, it will be assumed that greater task innovation will lead to higher users' satisfaction.

- *H10: Greater Task innovation positively influences Users' Satisfaction with portals*

## 2.5 Discussion and Conclusions

This paper aims at developing a theoretical framework based on Delone and Mclean IS Success Model (2003), which explains the success of Web portals in bringing benefits at the individual level in terms of improved task innovation. Indeed, organizations are investing money in establishing and running portals, without being aware of the benefits that can be gained on the individuals' level that justify these investments. In fact, portals offer the potential for substantially

١٥

improving task performance and innovation, since they have changed the way people handle information, communicate, share knowledge and perform tasks. But performance gains are often obstructed by users' unwillingness to accept and use available portals. Companies should be aware of the factors affecting users' intention to use portals, and their satisfaction, in order to guarantee that the investment made in establishing and running portals has achieved the intended benefits.

Web portals are considered as enablers for business intelligence since they are used to gather information existing in disparate databases throughout the company, and then display it on a single point of access. They help in the retrieval of accurate and timely information about financial operations, customers, and products, in order to gain analytical insight into business problems and opportunities, by this way decision makers and strategists are supported in the decision making process. Besides, portals play a role in the learning process, since they provide users with collaborative tools that fosters collaboration among internal and external parties, in addition to the possibility to gain access to all needed information that help in performing tasks and executing business processes; thus promoting learning opportunities, task performance and innovation.

This study has made a major contribution to theory, since a theoretical framework has been developed to be used in explaining the success of Web portals in promoting task innovation. The dimensions that affect the perceived quality of portals are identified, and the measures that affect each quality dimension are also defined. The developed theoretical framework explains that intention to use portals and users' satisfaction are affected by the level users perceive quality in terms of portals' system quality and information quality. Higher perceived quality will lead to higher satisfaction and higher intention to use and usage level. Positive experience of portals' use, measured in terms of improved task innovation, which in turn would lead to more satisfaction and more intention to continued usage. Another contribution is the identification of the main functions and features of portals (Content management and tailorability, Searchability, Collaboration, Integration, Security, Scalability, Accessibility and Integration), then the incorporation of these functions and features into the quality dimensions to which they contribute to. To clarify, portals' system quality can be measured by how much the portal can provide users with needed collaborative tools or search capabilities. Also the perceived portals' system quality is dependent on the degree of the portals' security and accessibility.

Practically, contributions gained by this study can be observed in terms of the benefits decision makers, strategists, operational employees and IT developers can gain. Assessing portals success in improving task innovation is important to help managers (i.e. decision makers) in making appropriate decisions concerning the adoption of portals' technology, by weighing its benefits against the costs needed to establish and run such a technology. Managers should also be aware of the factors affecting the intention of their employees to use portals' technology, in order to be able to provide them with whatever is needed to increase their usage level. Gaining access to needed useful information existing in disperse sources in a timely, efficient and effective manner makes it more possible for decision makers to make strategic and more informed decisions. From the perspective of information systems, assessing portals' success gives some insight to IT developers and designers concerning what aspects should be taken when designing and establishing high quality portals, and what functions and features should be contained that would affect the perceived quality of portals and therefore users' intention to use portals. IT designers should be able to translate users' expectations of the delivered quality of portals into appropriate quality specifications. At the operational level, portals act as a knowledge management system that promotes learning opportunities, by providing operational employees with useful information that help them in performing their job, and this may be reflected afterwards on the level of task innovation.

Although this study has made a contribution to theory by arriving at developing a theoretical framework that explains Web portals' success in promoting task innovation from the perspective of business intelligence, the developed model still needs to be validated. The proposed theoretical framework is more likely to be appropriate to be used and validated within the context of Web portals, as one type of knowledge management systems, but may not be valid for generalization to other contexts or other types of information systems.

In this study, "Service Quality" construct is assumed to be insignificant, since the quality of business intelligence systems, such as Web portals, depends mainly on the quality of the system, represented by the design characteristics and functioning capabilities, and the quality of information provided. However, if the Web portal success is to be measured in contexts other than business intelligence, we believe that "Service Quality" construct need to be included.

## References

1. Seddon, P. (1997) "A Re-specification and Extension of the DeLone and McLean Model of IS Success", *Information Systems Research*, Vol. 8, No. 3.
2. Petter, S., and McLean, E. (2009) " A meta-analytic assessment of the DeLone and McLean IS success model: An examination of IS success at the individual level", *Information & Management*, Vol. 46, pp.159–166.
3. Rai, A., Lang, S.S., and Welker, R.B. (2002), "Assessing the validity of IS Success Models: An empirical test and theoretical analysis". Information Systems Research, Vol. 13, pp. 50–69
4. Dias, C., (2001), "Corporate portals: a literature review of a new concept in Information Management", International Journal of Information Management, Vol. 21, pp. 269–287

5. Molla, A., and Licker, P. (2001), " E-Commerce Systems Success: An Attempt to extend and respecify the Delone and Mclean Model of IS Success", *Journal of Electronic Commerce Research*, Vol. 2, No. 4, pp. 131- 141

6. Seddon, P., Staples, S., Patnayakuni, R., and Bowtell, M. (1999), "Dimensions of Information Systems Success", Communications of the Association of Information Systems, Vol. 2

7. Yang, Z., Cai, S., Zhou, Z., and Zhou, N. (2005), "Development and validation of an instrument to measure user perceived service quality of information presenting Web portals", Information & Management, Vol. 42, pp. 575–589.

8. Moraga, A., Calero, C., and Piattini, M. (2006), "Comparing Different Quality Models for Portals", Online Information Review, Vol. 30, Iss: 5, pp.555 – 568.

9. Moraga, A., Calero, C., and Piattini, M. (2004), "A First Proposal of Portal Quality Model", IADIS International Conference e-Society

10. Manouselis, N., and Sampson, D. (2004) "Multiple Dimensions of User Satisfaction as Quality Criteria for Web Portals", IADIS International Conference WWW/Internet

11. Wang, Y., and Liao, Y. (2008), " Assessing eGovernment systems success: A validation of the DeLone and McLean model of information systems success", *Government Information Quarterly* 25, pp. 717–733.

12. Delone, W., and Mclean, E. (2003) "The DeLone and McLean Model of Information Systems Success: A Ten-Year Update", Journal of Management Information Systems, Vol. 19, No. 4, pp. 9–30.

13. Urbach, N., Smolnik, S., and Riempp, G. (2011), "Determining the improvement potentials of employee portals using a performance-based analysis", Business Process Management Journal, Vol. 17, No. 5, pp. 829-845

14. Lee, Y., Strong, D., Kahn, B., and Wang, R. (2002), "AIMQ: a methodology for information quality assessment", Information & Management, Vol. 40, pp. 133–146.

15. Masrek, M., Abdul Karim, N., and Hussein, R. (2007), "Investigating corporate intranet effectiveness: a conceptual framework", Information Management & Computer Security, Vol. 15 No. 3, pp. 168-183.

16. Urbach, N., Smolnik, S., and Riempp, G. (2010) " An empirical investigation of employee portal success", *Journal of Strategic Information Systems*, Vol. 19, pp. 184–206

17. Roses, L., Hoppen, N., and Henrique, J. (2009), "Management of perceptions of information technology service quality", Journal of Business Research, Vol. 62, pp. 876–882.

18. Wu, J., and Wang, Y. (2006) " Measuring KMS success: A respecification of the DeLone and McLean's model", *Information & Management*, Vol. 43, pp. 728–739.

19. Peter, S., DeLone, W., and McLean, E. (2008), "Measuring information systems success: models, dimensions, measures, and interrelationships", European Journal of Information Systems, Vol. 17, pp. 236–263.

20. Torkzadeh, G., and Doll, W.J. (1999), "The development of a tool for measuring the perceived impact of information technology on work", *Omega, International Journal of Management Science*, Vol. 27, pp. 327- 339.

21. Seddon, P., and Kiew, M. (1996),"A Partial Test and Development of DeLone and McLean's Model of IS Success ", Australasian Journal of Information Systems, Vol. 4, No. 1, pp. 90-108

22. Igbaria, M., and Tan, M. (1997), "The consequences of Information Technology Acceptance on Subsequent Individual Performance", Information & Management, Vol. 32, pp. 113-121.

23. McGill, T., Hobbs, V., and Klobas, J. (2003), " User-Developed Applications and Information Systems Success: A Test of DeLone and McLean's Model", Information Resources Management Journal, Vol. 16, pp. 24-45.

24. Garrity, E., Glassberg, B., Kim, Y., Sanders, G.L., and Shin, S. (2005), "An experimental investigation of Web-based information systems success in the context of electronic commerce", Decision Support Systems, Vol. 39, pp. 485– 503

25. Sugianto, L.F., and Tojib, D. (2006), "Modeling user satisfaction with an employee portal", *International Journal of Business*, Vol. 24, pp. 339–348.

26. Rahim, M. M., and Singh, M. (2006), "Understanding benefits and Impediments of B2E E-Business Systems Adoption: Experience of Two Large Australian Universities", IADIS International Conference e-Society.

27. Raol, J., Koong, K., Liu, L., and Yu, C. (2003), "Identification and Classification of Enterprise Portal Functions and Features", Industrial Management and Data Systems, 103/9, pp. 693-702.

28. Benbya, H., Passiante, G., and Belbaly, N. (2004), "Corporate portal: a tool for knowledge management synchronization", International Journal of Information Management, Vol. 24, pp. 201–220

29. Detlor, B. (2000), " The corporate portal as information infrastructure: Towards a framework for portal design", International Journal of Information Management, Vol. 20, pp. 91-101.

30. Rahim, M.M. (2008), "A Qualitative Evaluation of an Instrument for Measuring the Influence of Factors Affecting Use of Business-to-Employee (B2E) Portals", *World Academy of Science, Engineering and Technology*, Vol. 46.

31. Urbach, N., Smolnik, S., and Riempp, G. (2009) "A Conceptual Model for Measuring the Effectiveness of Employee Portals", *Proceedings of the Fifteenth Americas Conference on Information Systems, San Francisco, California*, August 6th-9th 2009

# الملخص

تهدف هذه الدراسة بشكل أساسي إلى تطوير إطار نظري يفسر مدى نجاح البوابات الالكترونية في تحفيز الإبداعية في العمل. ويعتبر هذا مهماً حيث أن هناك القليل من الدراسات التي تطرقت لدراسة ذلك من منظور الذكاء الوظيفي. اعتمدت هذه الدراسة على نموذج ديلون وماكليين الذي استخدم لتفسير نجاح النظم المعلوماتية، حيث اعتبر هذا النموذج كقاعدة نموذجية لتطوير نموذج الدراسة الحالية. في هذه الدراسة، تم تحديد وتعريف متغيري جودة النظام وجودة المعلومات بالاستناد إلى خصائص البوابات الالكترونية حيث تم ربط أهم وأبرز ملامح وميزات البوابات الإلكترونية بالمتغيرات التي تحدد جودة النظام وجودة المعلومات التي نحصل عليها باستخدام البوابات الالكترونية. كما يعتبر النموذج الذي تم تطويره في هذه الدراسة مفيداً من الناحيتين النظرية والعملية. نظرياً، تم تحديد الأبعاد التي تؤثر على جودة البوابات الإلكترونية كما يدركها المستخدمون، بالإضافة إلى تحديد المقاييس والمعايير التي تؤثر على كل بعد من أبعاد الجودة. من الناحية العملية، هناك عدة فوائد من الممكن أن يكتسبها كل من أصحاب القرار والاستراتيجيين والموظفين ومطوري النظم المعلوماتية، حيث أن تقييم مدى نجاح البوابات الالكترونية في تحسين الإبداعية في العمل يساعد المدراء وأصحاب القرار على اتخاذ القرارات المناسبة فيما يتعلق بتبني تقنية البوابات الالكترونية ، بعد أن يتم مقارنة النتائج المرجوة من استخدام مثل هذه التقنية وتكلفة إنشائها وتطبيقها. إضافةً إلى ذلك، تقييم مدى نجاح البوابات الالكترونية يمكن مطوري ومصممي النظم المعلوماتية من تحديد الجوانب التي يجب التركيز عليها عند تصميم وتطوير بوابات إلكترونية ذات جودة عالية تحتوي كافة الأدوات والوسائل والخصائص التي تؤثر على الجودة، مما يمكن مستخدمي البوابات على لمس جودتها ويجعلهم أكثر استعداداً لاستخدامها.